\documentclass[cits]{PoS}
\pdfoutput=1

\usepackage{amsmath,graphicx}

\DeclareMathOperator{\tr}{tr}
\DeclareMathOperator{\re}{Re}
\DeclareMathOperator{\im}{Im}
\newcommand{\hxi}{\hat{\xi}}

\title{Distributions of individual Dirac eigenvalues\\ for QCD at
  non-zero chemical potential:\\ RMT predictions and lattice results}

\ShortTitle{Distributions of individual Dirac eigenvalues for QCD at
  non-zero chemical potential}

\author{Gernot Akemann$^a$, Jacques Bloch$^b$, \speaker{Leonid
    Shifrin}$^a$, and Tilo Wettig$^b$\\
  $^a$Department of Mathematical Sciences \& BURSt Research Centre,\\
  \phantom{$^a$}Brunel University West London, Uxbridge UB8 3PH,
  United Kingdom\\ 
  $^b$Institute for Theoretical Physics, University of Regensburg, 
  93040 Regensburg, Germany\\
  E-mail: \email{Leonid.Shifrin@brunel.ac.uk}}

\abstract{For QCD at non-zero chemical potential $\mu$, the Dirac
  eigenvalues are scattered in the complex plane. We define a notion
  of ordering for individual eigenvalues in this case and derive the
  distributions of individual eigenvalues from random matrix theory
  (RMT).  We distinguish two cases depending on the parameter
  $\alpha=\mu^2F^2V$, where $V$ is the volume and $F$ is the familiar
  low-energy constant of chiral perturbation theory.  For small
  $\alpha$, we use a Fredholm determinant expansion and observe that
  already the first few terms give an excellent approximation.  For
  large $\alpha$, all spectral correlations are rotationally
  invariant, and exact results can be derived.  We compare the RMT
  predictions to lattice data and in both cases find excellent
  agreement in the topological sectors $\nu=0,1,2$.}

\FullConference{The XXV International Symposium on Lattice Field Theory\\
		July 30 - August 4, 2007\\
                Regensburg, Germany}

\begin{document}

\section{Introduction}

Studies of the properties of the Dirac operator spectrum in gauge
theories, including QCD, have a long history. For example, the
low-lying Dirac modes provide information about spontaneous chiral
symmetry breaking through the Banks-Casher relation.  The Dirac
operator spectrum is also a natural object to study in lattice QCD. In
the deep infrared, QCD in the $\epsilon$-regime can be described by
the chiral random matrix theory (RMT) introduced in Ref.~\cite{RMT}.
One of the advantages of RMT is that many exact analytical results can
be derived.  These results contain the low-energy constant (LEC)
$\Sigma$ of chiral perturbation theory (chPT), and in some cases also
the LEC $F$.  These LECs can then be determined by fitting lattice
data to the RMT curves.

The observables that are most natural to compute in RMT are the
spectral density correlation functions. At zero chemical potential
$\mu$, all of them are known in RMT \cite{RMT,ADMN}. On the other
hand, one can consider individual Dirac eigenvalue distributions
(IED). From the lattice QCD point of view, these are the most natural
observables to measure directly. Certain quantities such as the
average positions of the eigenvalues are more pronounced in IEDs and
therefore require less statistics to be measured reliably.

At $\mu=0$, the Dirac operator is anti-Hermitian and has a purely
imaginary spectrum. In this case, all IEDs are known analytically in
RMT \cite{pk_real}. They have become a standard tool in lattice QCD to
extract $\Sigma$ in sectors of fixed topology. At $\mu\ne0$, the Dirac
operator is no longer anti-Hermitian, and its eigenvalues are
scattered in the complex plane.  Our work is based on an RMT for
$\mu\ne0$ \cite{James} which has an eigenvalue representation and for
which all complex density correlations (both quenched and unquenched)
have been calculated. The same results can be obtained from the RMT
for $\mu\ne0$ introduced earlier by Stephanov \cite{Misha} or from
chPT in the $\epsilon$-regime \cite{SplitV,BA} and are universal in
that sense. These results have been compared to data from quenched
lattice simulations with staggered \cite{AOW} and overlap
\cite{Bloch:2006cd} fermions at $\mu\neq0$.  A virtue of $\mu\neq0$ is
that it couples to $F$ in leading order of chPT \cite{Toublan:1999hx}
so that a comparison with lattice data allows us to extract $F$
\cite{AOW}.

For the IEDs at $\mu\ne 0$ much less is known. One of the problems
here is to define an ordering of complex eigenvalues. Previous work on
the repulsion between complex eigenvalues in RMT \cite{GHS88} and on
the lattice \cite{Markum:1999yr} was done in the bulk of the spectrum,
where no link to chPT is apparent. In the present work, we are
interested in IEDs for eigenvalues close to the origin since they
provide information on topological properties and LECs.  

We first define the general notion of IEDs for complex eigenvalues,
and then compute the first few IEDs approximately by truncating a
so-called Fredholm determinant expansion to the first few terms. It
was already observed in Ref.~\cite{Akemann:2003tv} for $\mu = 0$ that
this is a very good approximation.  For large values of the parameter
$\alpha=\mu^2F^2V$, where $V$ is the volume, we are able to derive all
IEDs in closed form \cite{AS}. We use these exact results as a
consistency check of the Fredholm determinant expansion. Our results
are then compared to the lattice data of Ref.~\cite{Bloch:2006cd}, in
which the overlap Dirac operator for $\mu\ne 0$ was constructed. This
operator has good chiral properties (it satisfies a Ginsparg-Wilson
relation, has an exact lattice chiral symmetry and exact zero modes,
and satisfies the index theorem), which is essential for the present
work.  The same operator is obtained if a chemical potential is
introduced in the domain-wall fermion formalism in the limit of
infinite extent of the fifth dimension \cite{Bloch:2007xi}.  Due to
the sign problem at $\mu\ne 0$, the lattice analysis in
Ref.~\cite{Bloch:2006cd} was restricted to the quenched case, and this
restriction on the lattice data applies to this work as well.

\section{Individual eigenvalue distributions for complex eigenvalues} 

Consider an operator with a finite number $N$ of complex eigenvalues,
distributed according to a joint probability distribution
${\cal{P}}(\{z\})$ which is symmetric in all its arguments. We also
assume a $z\to -z$ symmetry and only consider the upper half-plane
${\mathbb{C}}_+$.  The partition function is then given by $Z =
\int_{{\mathbb{C}}_+} \prod_{j=1}^N d^2z_j {\cal{P}}(\{z\})$, and
the spectral density correlation functions are defined as
\begin{equation}
  \label{Rkdef}
  R_k(z_1,...,z_k)=\frac1Z\frac{N!}{(N-k)!}
  \int_{{\mathbb{C}}_+} d^2 z_{k+1}\ldots d^2 z_N\,
  {\cal{P}}_N(z_1,...,z_N)\:.
\end{equation} 
Now consider any one-parameter family of mutually non-intersecting
closed contours $\partial J[\eta]$ which cover ${\mathbb{C}}_+$. For
fixed $\eta$, $\partial J[\eta]$ is the boundary of a set $J[\eta]$.
Let us parametrize the contour as $\partial J(z(\tau))$ with
$z(\tau)\equiv x(\tau)+iy(\tau)$.  We then define the $k$-th
eigenvalue distribution $p_k(J,\tau)$ as the probability that $k-1$
eigenvalues are inside $J$, one is at the point $z(\tau)$ on the
boundary $\partial J$, and the remaining $N-k$ are in the complement
$\bar{J}$,
\begin{equation}
  \label{pdef}
  p_k(J,\tau)\equiv \frac{k}{Z}
  \binom{N}{k}\prod_{j=1}^{k-1}\int_{J} d^2z_j
  \prod_{i=k+1}^N\int_{\bar J} d^2z_i\,{\cal P}(\{z\})
  \big|_{z_k=z(\tau)}\:.
\end{equation} 
Note that the eigenvalue ordering is induced by the entire contour
family. The $\{\eta,\tau\}$ play the role of generalized polar
coordinates. In the following, the argument $J$ of $p_k$ will be
suppressed. It is possible \cite{AS} to express all $p_k(\tau)$
through the densities Eq.~\eqref{Rkdef}. In particular, for the
distribution of the first eigenvalue one obtains
\begin{equation}
  p_1(\tau) = R_1(z(\tau))-\int_J d^2z_1 \, R_2(z_1,z(\tau))
  +\frac{(-1)^2}{2!}\int_Jd^2z_1\int_Jd^2z_2\, R_3(z_1,z_2,z(\tau))+\ldots
  \label{p1exp}
\end{equation}
One can show that the integrated distributions $P_k(\eta)\equiv\int
d\tau\, j(\eta,\tau)p_k(\eta,\tau)$ are normalized as
$\int_{\eta_\text{min}}^{\eta_\text{max}} P_k(\eta)d\eta = 1$, where
$[\eta_\text{min},{\eta_\text{max}}]$ is the range of $\eta$, and $
j(\eta,\tau) = |(\partial x, \partial y)/(\partial \eta,
\partial \tau)|$ is the Jacobian of the transformation from $(x,y)$ to
$(\eta,\tau)$.  The proof is similar to that for real eigenvalues
\cite{Akemann:2003tv}.  We emphasize that in this framework the choice
of the contour family becomes part of the definition of the quantities
we measure (i.e., the $p_k$'s). Different contour families in general
lead to different $p_k$'s. However, one relation always holds
trivially, namely $\sum_{k=1}^{N} p_k(z) = R_1(z)$.

\section{RMT predictions}
\label{rmt}

The partition function of the matrix model we use \cite{James} reads
\begin{equation}
  \label{model}
  Z_{\nu} = \int dA dB\, \exp\{-N\tr(AA^\dagger+BB^\dagger)\}
  \prod_{f=1}^{N_f}\det 
  \left(\begin{array}{cc} 
      m_f & iA+\hat\mu B\\
      iA^{\dagger}+\hat\mu B^{\dagger} & m_f 
    \end{array}\right).
\end{equation}
Here, $A$ and $B$ are complex $(N+\nu)\times N$ matrices with no
further symmetries, $\nu\geq0$ is the topological charge, the $m_f$
are the masses of $N_f$ flavors of dynamical quarks, and $\hat\mu$ is
the chemical potential in the matrix model.  In the large-$N$ limit
this model describes QCD in the $\epsilon$-regime.  All density
correlation functions of this model follow from the kernel
$K_{N}(z_i,z^*_j)$ of bi-orthogonal polynomials with respect to the
weight
\begin{equation}
  w^{(N_f,\nu)}(z_j)=\prod_{f=1}^{N_f}m_f^\nu (m_f^2-z_j^2)
  |z_j|^{2\nu+2}K_\nu\left(\frac{N(1+{\hat\mu}^2)}{2{\hat\mu}^2}|z_j|^2\right)
  e^{\frac{N({\hat\mu}^2-1)}{4{\hat\mu}^2}\left(z_j^2+z_j^{*\,2}\right)}\:,
\end{equation}
where $K_\nu$ (and $I_\nu$ below) are modified Bessel functions,
according to
\begin{equation}
  \label{Rksol}
  R_k(z_1,\ldots,z_k)
  =\prod_{\ell=1}^k w^{(N_f,\nu)}(z_\ell)
  \det_{1 \leq i,j \leq k}K_{N}(z_i,z^*_j)
  =:\det_{1 \leq i,j \leq k}\mathcal K_{N}(z_i,z^*_j)\:.
\end{equation}
We rescale the parameters of the model such that the parameters
$\alpha\equiv 2N\hat\mu^2\,(=VF^2\mu^2)$, $\eta_f\equiv
Nm_f\,(=V\Sigma m_f)$, and $\xi_k\equiv Nz_k\,(=V\Sigma z_k)$ stay
finite in the large-$N$ limit.  The scaling of these parameters in
terms of the LECs of chPT is given in parentheses. In the quenched
case, the microscopic kernel ${\cal K}_s(z_i,z_j^*) =
\lim_{N\to\infty}{\cal K}_N(z_i/N,z_j^*/N)/N$ is given by \cite{James}
\begin{equation}
  \label{kernel}
  {\cal{K}}_s(z_i,z_j^*) = \frac{|z_iz_j^*|^{\nu+1}}
  {2\pi\alpha(z_iz_j^*)^\nu}
  \sqrt{K_{\nu}\left(\frac{|z_i|^2}{4\alpha}\right)
    K_{\nu}\left(\frac{|z_j^*|^2}{4\alpha}\right)}
  e^{-\frac{\re(z_i^2+{z_j^*}^2)}{8\alpha}}\int_0^1 dt\,
  e^{-2\alpha t}I_{\nu}(z_i\sqrt{t})I_{\nu}(z_j^*\sqrt{t})\:,
\end{equation}
and the microscopic spectral density follows as
$\rho_1(\xi)={\cal{K}}_s(\xi,\xi^*)$.

\subsection*{Approximate computations for arbitrary $\alpha$}

\begin{figure}[-t]
  \centering
  \hspace*{-3mm} \includegraphics[width=50mm]{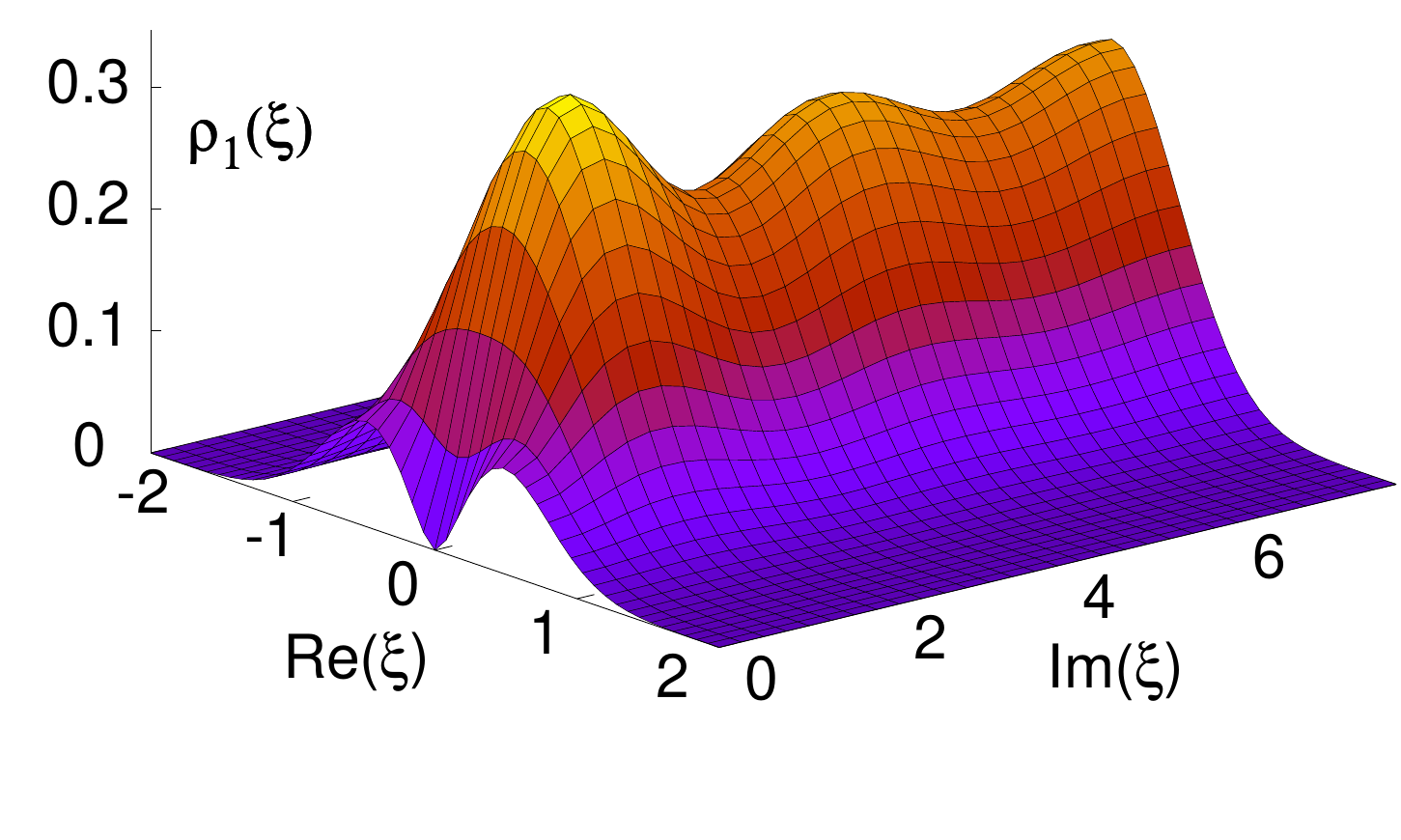} \hspace*{-1mm}
  \includegraphics[width=50mm]{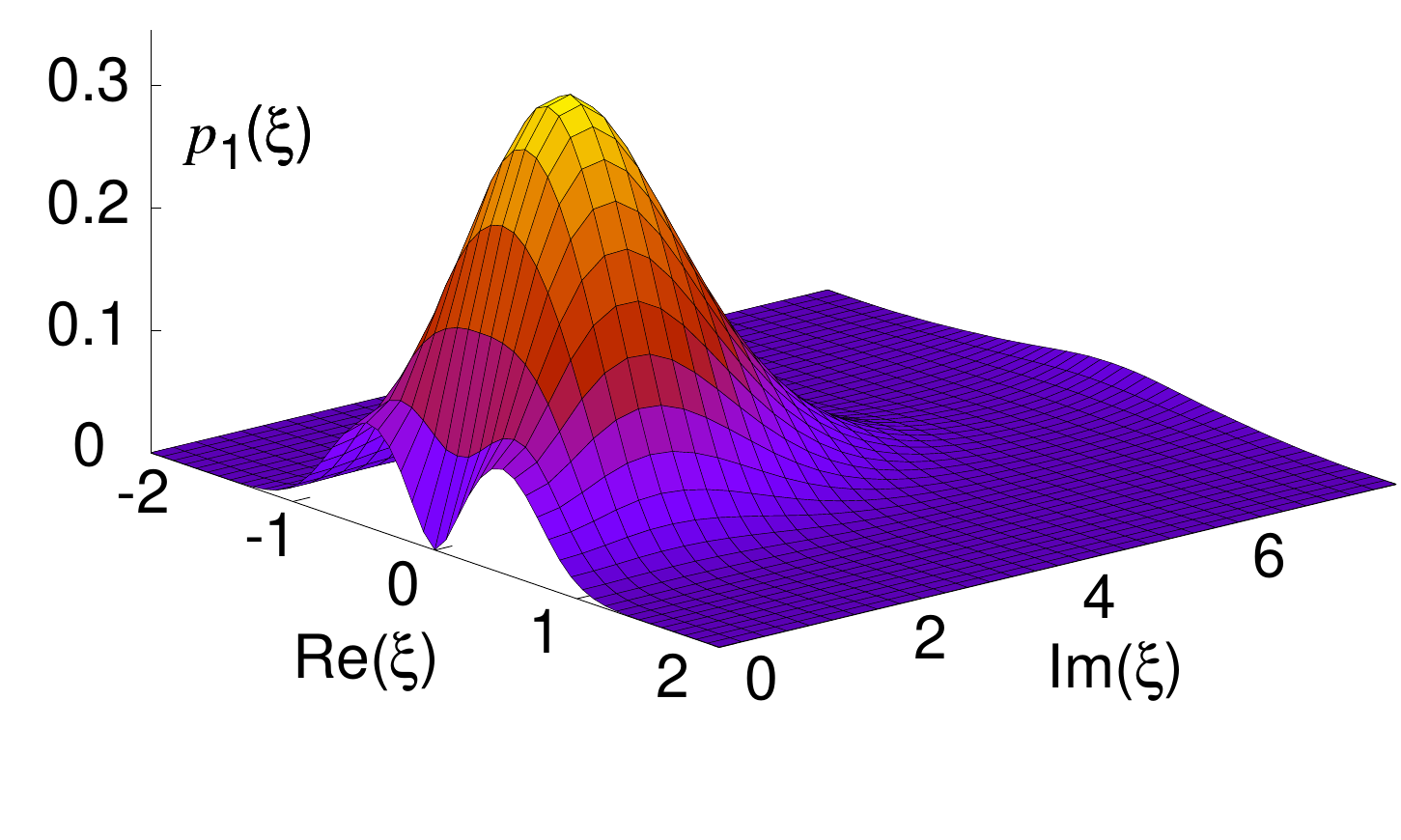} \hspace*{-1mm}
  \includegraphics[width=50mm]{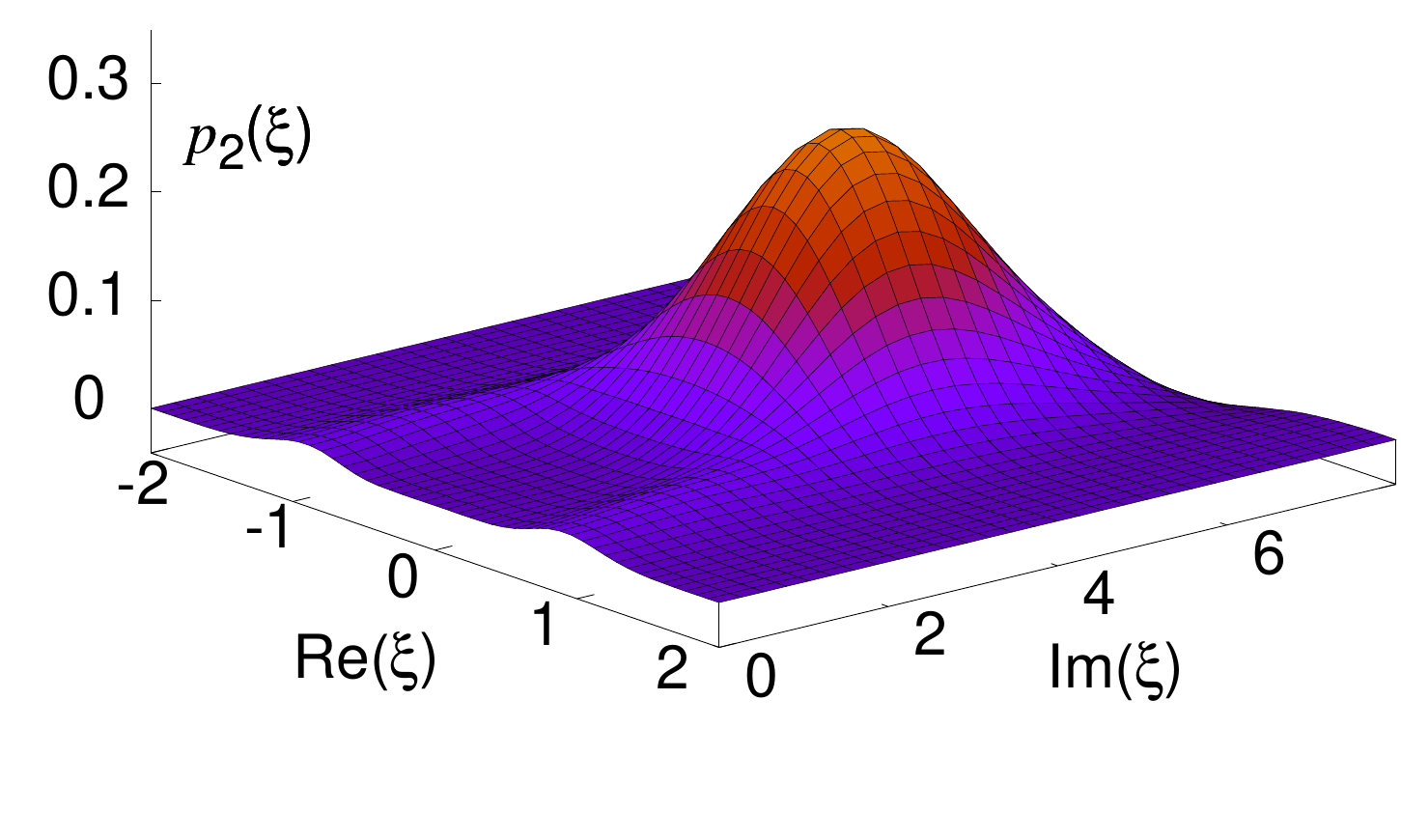} \\
  \includegraphics[width=48mm]{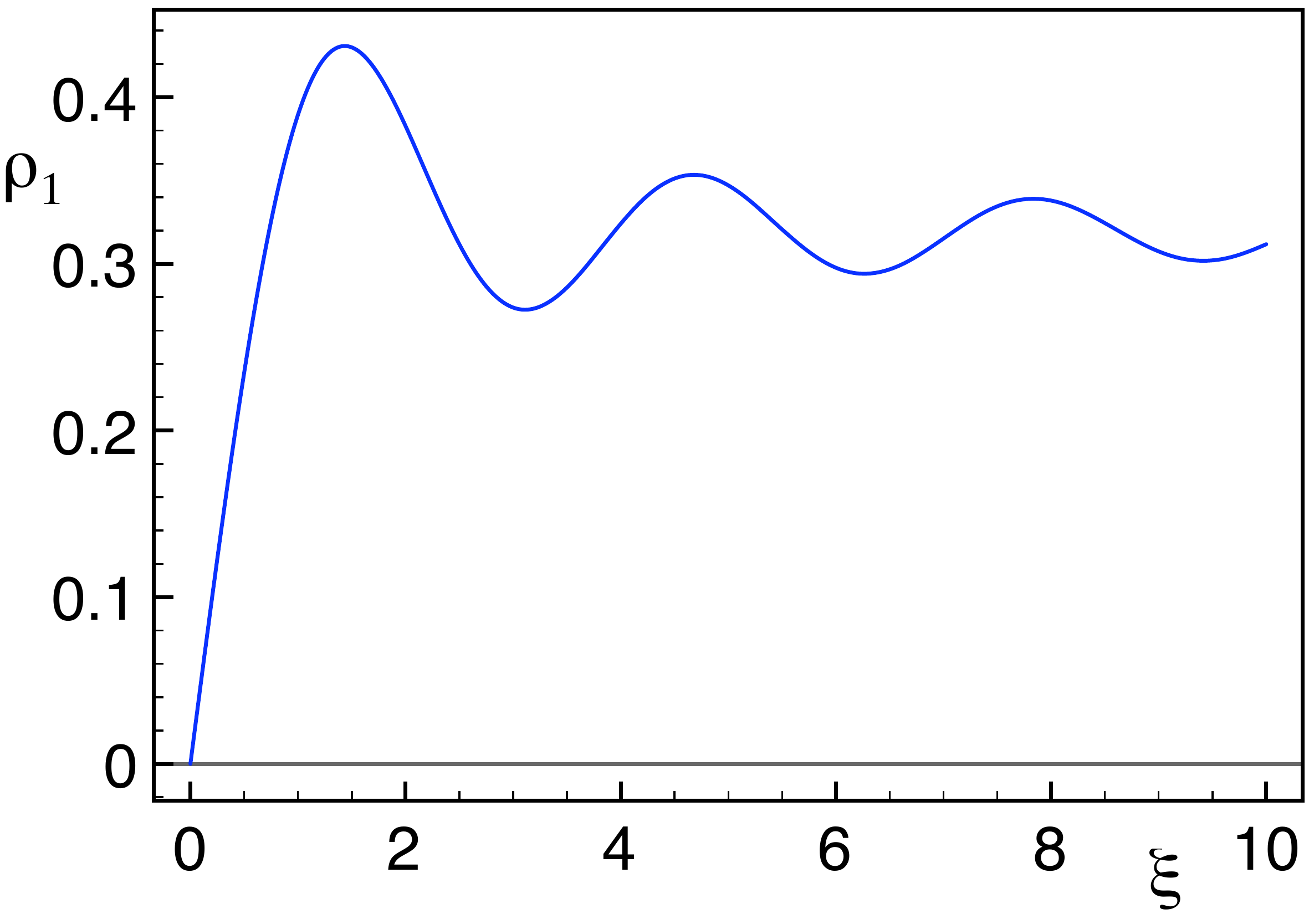} \hspace*{1mm}
  \includegraphics[width=48mm]{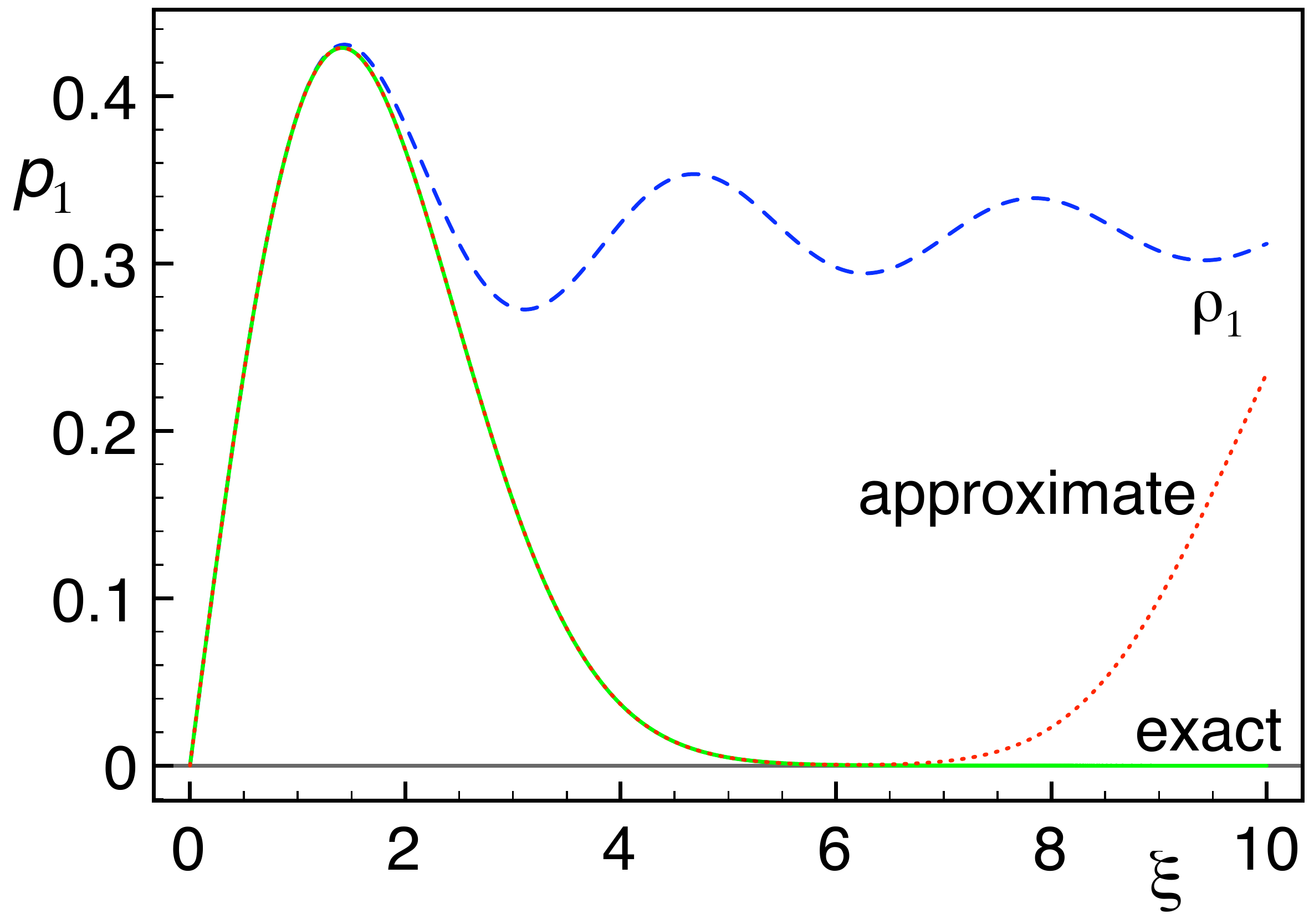} \hspace*{1mm}
  \includegraphics[width=48mm]{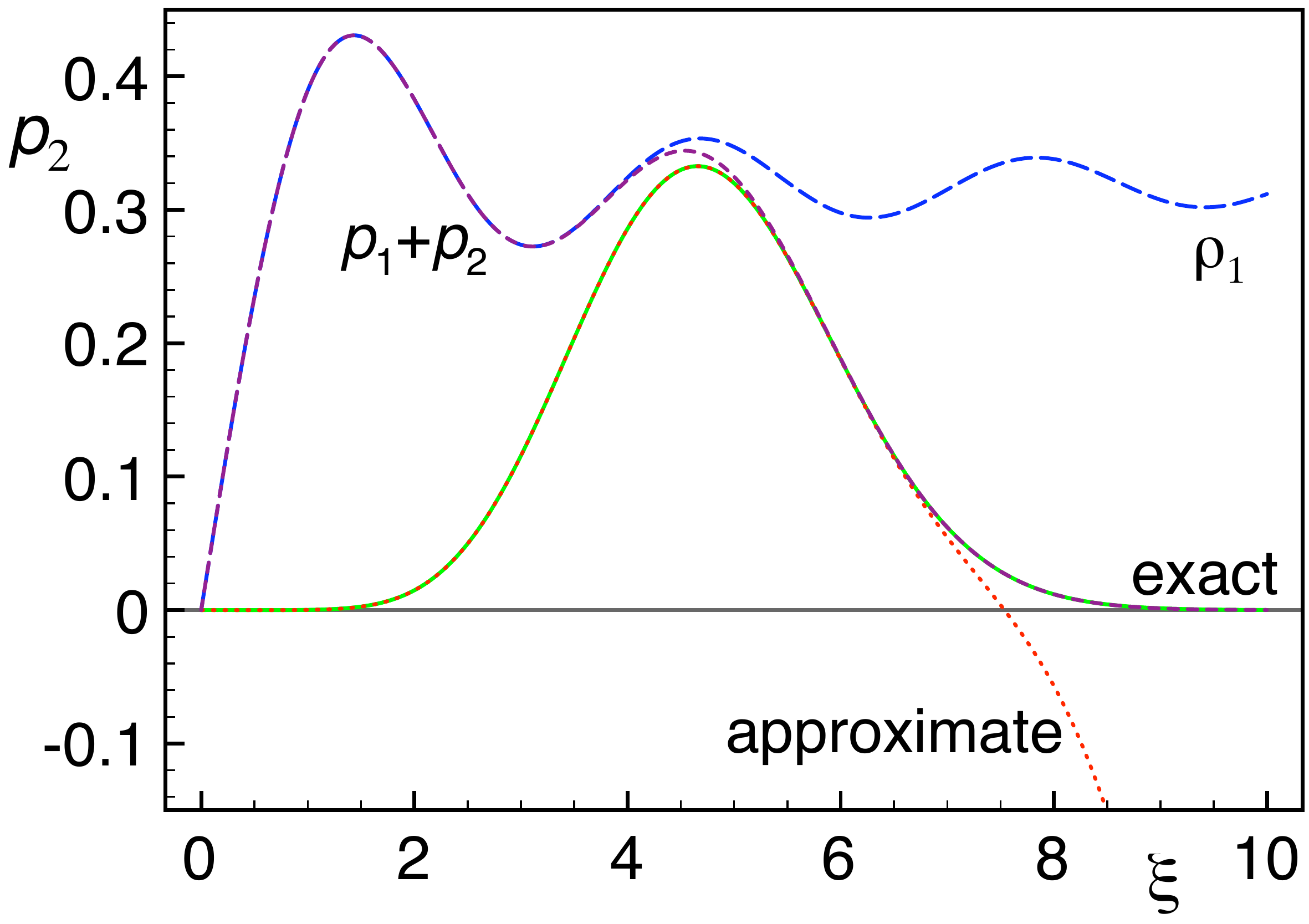}
  \caption{Top: 
    Microscopic density $\rho_1(\xi)$ (left) and distributions
    $p_1(\xi)$ (middle) and $p_2(\xi)$ (right) of the first and second
    eigenvalue for an elliptic family of $\partial J[\eta]$
    parametrized as $(\re\xi/5)^2+(\im\xi)^2 = \eta^2$ on $\mathbb
    C_+$, all for $\nu=0$ and $\alpha=0.174$.  
    Bottom: 
    Counterparts for $\mu = 0$.
    Left: Density.
    Middle: Density (dashed/blue), exact $p_1(\xi)$ (solid/green), and
    approximate $p_1(\xi)$ (dotted/red).
    Right: Density (long dashes/blue), exact $p_1(\xi)+p_2(\xi)$
    (short dashes/purple), exact $p_2(\xi)$ (solid/green), and
    approximate $p_2(\xi)$ (dotted/red).
    The deviations of the approximate $p_1$ and $p_2$ from the exact
    results are an artifact of the truncation of the Fredholm
    determinant expansion.} 
  \label{3d}
\end{figure}

In order to obtain the distribution $p_1$, we use Eq.~\eqref{p1exp}
and substitute the densities from Eqs.~\eqref{Rksol} and
\eqref{kernel}.  Similar formulas can be written down for other
$p_k$'s.  For practical purposes, we truncate the (so-called Fredholm
determinant) expansion in Eq.~\eqref{p1exp} to the first three terms
and the corresponding expansion for $p_2$ to the first two terms. The
density $\rho_1(\xi)$ and the distributions $p_1(\xi)$ and $p_2(\xi)$
of the first and second eigenvalue are shown in Fig.~\ref{3d},
together with their counterparts in the $\mu = 0$ case.  For $\mu = 0$
exact results are available \cite{pk_real} which facilitate a detailed
comparison. As for $\mu = 0$ \cite{Akemann:2003tv}, we see that the
expansion converges rapidly.  Higher-order terms merely assure that
$p_k(\xi)$ remains zero for large $|\xi|$.

\begin{figure}[-t]
  \centering 
  \includegraphics[width=48mm]{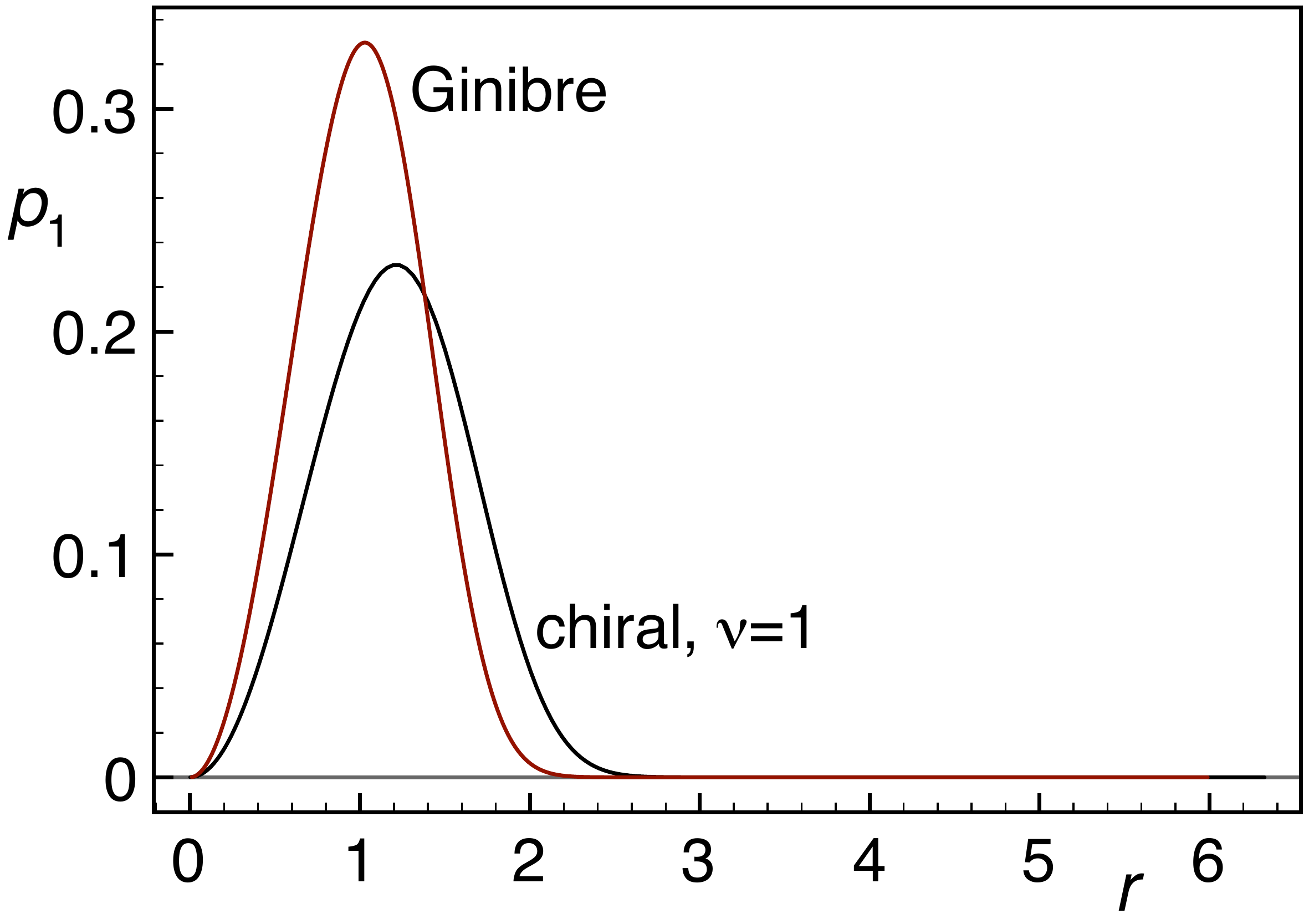}\hspace*{2mm}
  \includegraphics[width=48mm]{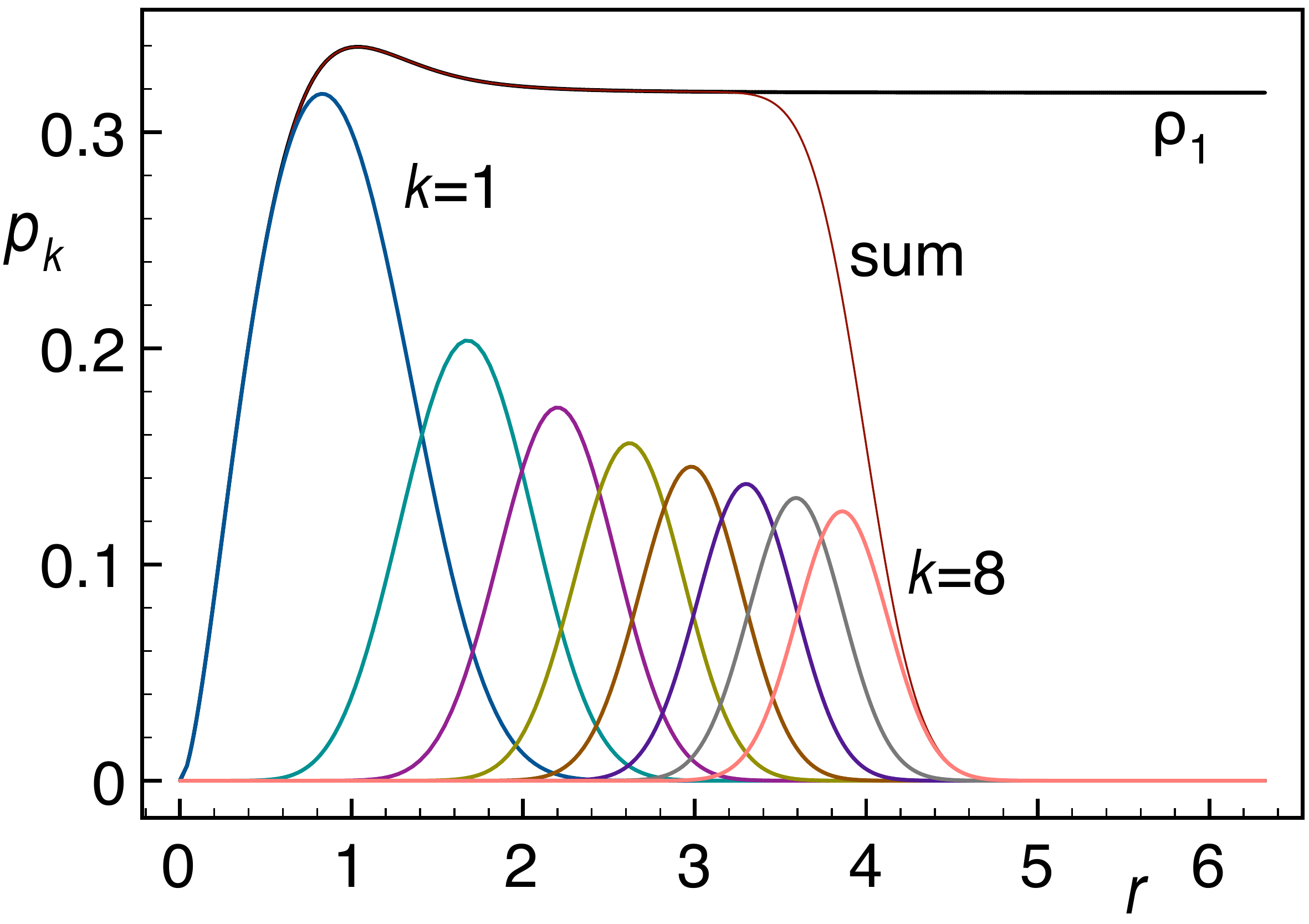}\hspace*{2mm} 
  \includegraphics[width=48mm]{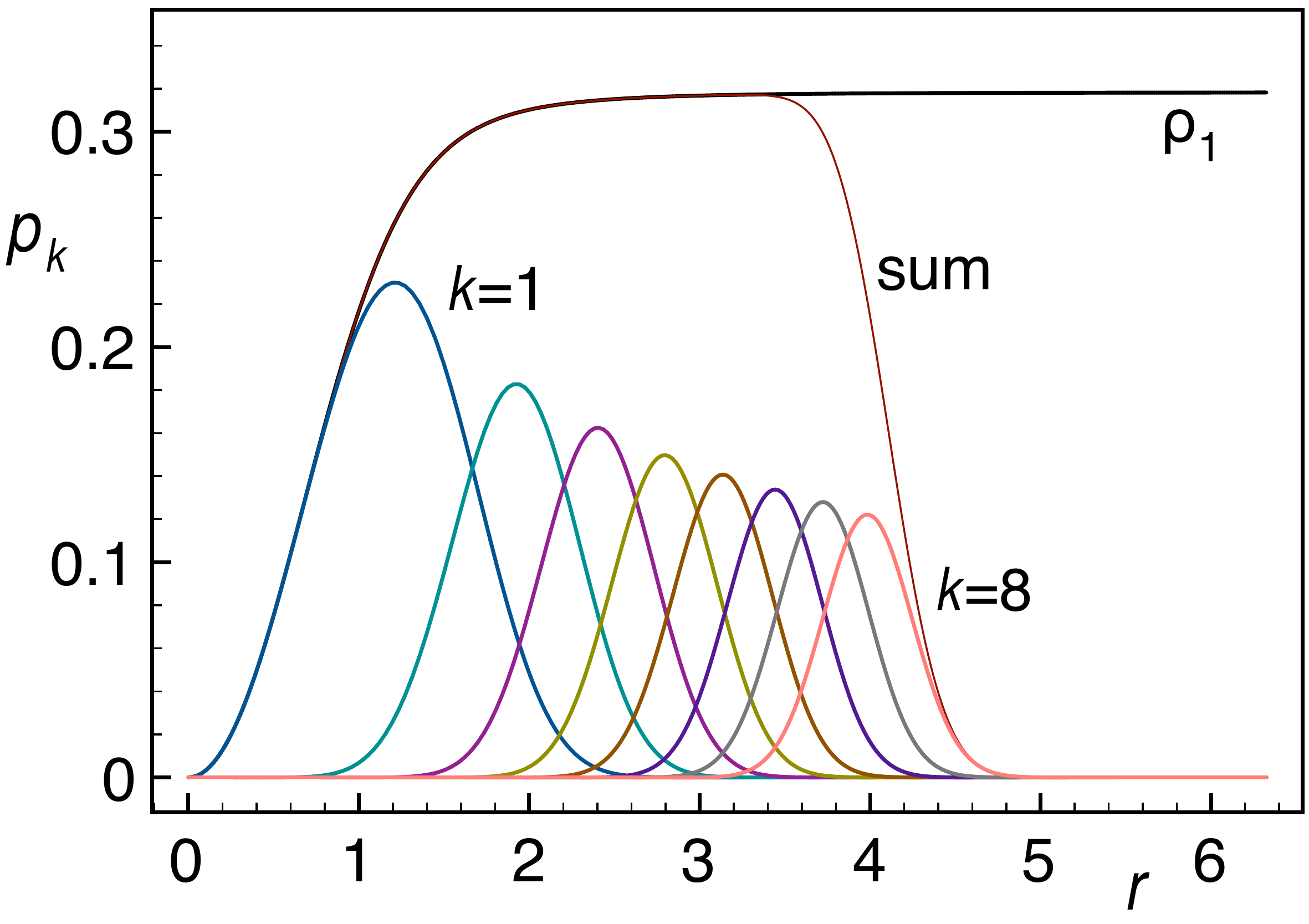}
  \caption{Left: First eigenvalue distribution as a function of the
    radius, for the Ginibre ensemble and for the $\nu = 1$ chiral
    ensemble in the $\alpha\to\infty$ limit from
    Eq.~\protect\eqref{p1}. The $\nu=1$ sector was chosen since for
    the comparison we need exactly one eigenvalue at the origin in
    both cases.  Middle: Spectral density Eq.~\protect\eqref{R1rad}
    and distributions of the first eight eigenvalues from
    Eq.~\protect\eqref{p1} (and similar for $p_k(r)$ with $k\ge 2$) as
    well as their sum, all in the $\alpha\to\infty$ limit and for
    $\nu=0$. Right: Same for $\nu = 1$.}
  \label{psum}
\end{figure}

\subsection*{Exact results in the large-$\alpha$ limit}

In the $\alpha\to\infty$ limit, the problem becomes radially
symmetric. For finite but large $\alpha$, the symmetry is apparent
close to the origin.  The limiting microscopic spectral density
expressed in the new variable $\hat{\xi}\equiv\xi/2\sqrt\alpha$ reads
\begin{equation}
  \lim_{\alpha\to\infty}  \rho_1(\hxi) =
  \frac{2|\hxi|^2}{\pi} K_\nu(|\hxi|^2)I_\nu(|\hxi|^2)\:.
  \label{R1rad}
\end{equation}
In this limit, we can derive a closed expression for all eigenvalue
distributions \cite{AS}.  Because of the rotational symmetry we choose
$\partial J$ to be a semi-circle in ${\mathbb{C}}_+$ of radius
$r\equiv|\hat{\xi}|$ and obtain for $p_1(r)$ 
\begin{equation}
  \label{p1}
  p_1(r)=-\frac{1}{\pi r}\frac{\partial}{\partial r}\prod_{\ell=0}^{\infty}
  \biggl\{\frac{r^{4\ell+2\nu+2} K_{\nu+1}(r^2)}{2^{2\ell+\nu}\ell!(\ell+\nu)!}
  + r^2\left[ K_{\nu+1}(r^2)I_{\nu+2}^{[\ell-2]}(r^2)    
  +    K_{\nu+2}(r^2)I_{\nu+1}^{[\ell-1]}(r^2) \right]
  \biggr\} \:.
\end{equation}
Here, we have introduced the incomplete Bessel function
$I_{\nu}^{[\ell]}(x)\equiv\sum_{n=0}^{\ell}(x/2)^{2n+\nu}/n!(n+\nu)!$
for $\ell\ge 0$, and zero otherwise.  Our result is analogous to the
result for the nearest-neighbor spacing distribution of the Ginibre
ensemble \cite{GHS88,Markum:1999yr}, which can be interpreted as the
distribution of the smallest nonzero eigenvalue if one eigenvalue is
fixed at zero.  Expressions for $p_k(r)$ with $k\ge 2$ are also
available \cite{AS,ABSW}. Fig.~\ref{psum} shows several $p_k$, summing
up nicely to the spectral density Eq.~\eqref{R1rad}.

\section{Lattice calculations}
\label{lat}

The lattice part of our work is based on the data obtained in
Ref.~\cite{Bloch:2006cd}. The overlap Dirac operator introduced there
is
\begin{equation}
  D_\text{ov}(\mu) = 1+\gamma_5\,\epsilon(\gamma_5 D_W(\mu))\:,
\label{D_ov}
\end{equation}
where $\epsilon$ is the sign function of a non-Hermitian matrix and
$D_W(\mu)$ is the Wilson Dirac operator at $\mu\ne 0$. This overlap
operator was shown to satisfy a Ginsparg-Wilson relation and to have
good chiral properties \cite{Bloch:2006cd,Bloch:2007xi}.
Equation~\eqref{D_ov} reduces to the standard overlap operator
\cite{overlap} at $\mu = 0$.

From the computational standpoint, the most demanding part is the
computation of the matrix sign function. For the present set of data,
this was done exactly using the spectral definition of the sign
function.  The lattice size is only $V = 4^4$, since high statistics
are needed for a comparison with RMT. The coupling in the Wilson
action is $\beta = 5.1$ in order to stay in the $\epsilon$-regime
(where RMT applies) for the first eigenvalue(s) \cite{Bloch:2006cd}.
The Wilson mass is $m_Wa = -2$ ($a$ is the lattice spacing), and the
quark mass is zero. Data were sampled in the topological sectors
$\nu=0$, 1, 2 for the values of $\mu a=0.1$, 0.2, 0.3, and 1.0,
corresponding to $\alpha=0.174$, 0.615, 1.42, and 4.51, respectively.
The number of configurations varied from about 9000 for $\mu a= 0.1$
to about $3000$ for $\mu a= 1.0$. The parameters $\Sigma$ and $F$ were
determined by a fit to the spectral density from Eq.~\eqref{kernel}.
For $\mu a= 1.0$, the data showed rotational invariance up to
$\hat{|\xi|} = 0.7$, and hence only the combination $\Sigma/F$ could
be determined by a fit to Eq.~\eqref{R1rad}. Our comparisons use these
values and are thus parameter-free.

\begin{figure}[-t]
  \centering 
  \includegraphics[width=48mm]{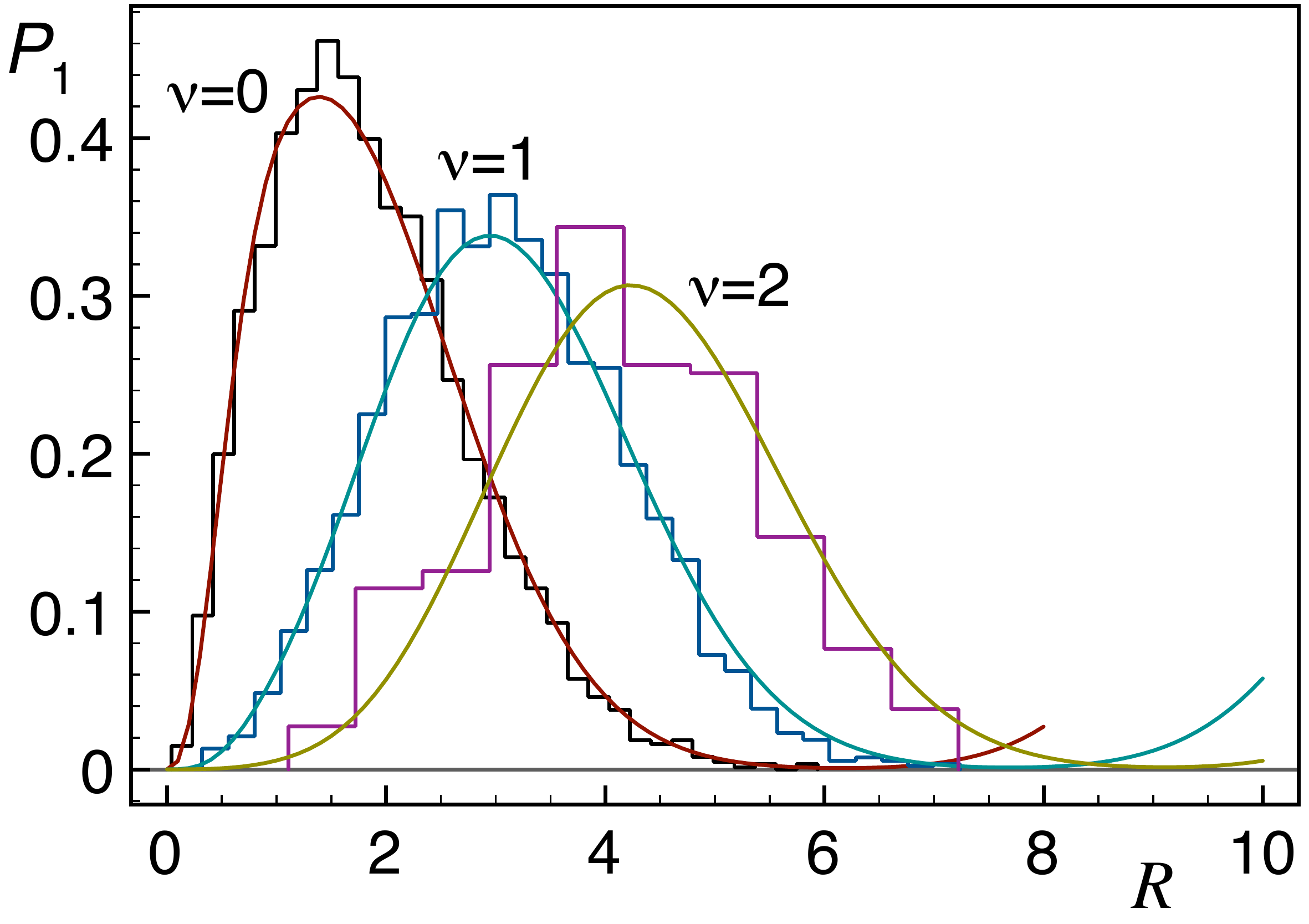}\hspace*{2mm} 
  \includegraphics[width=48mm]{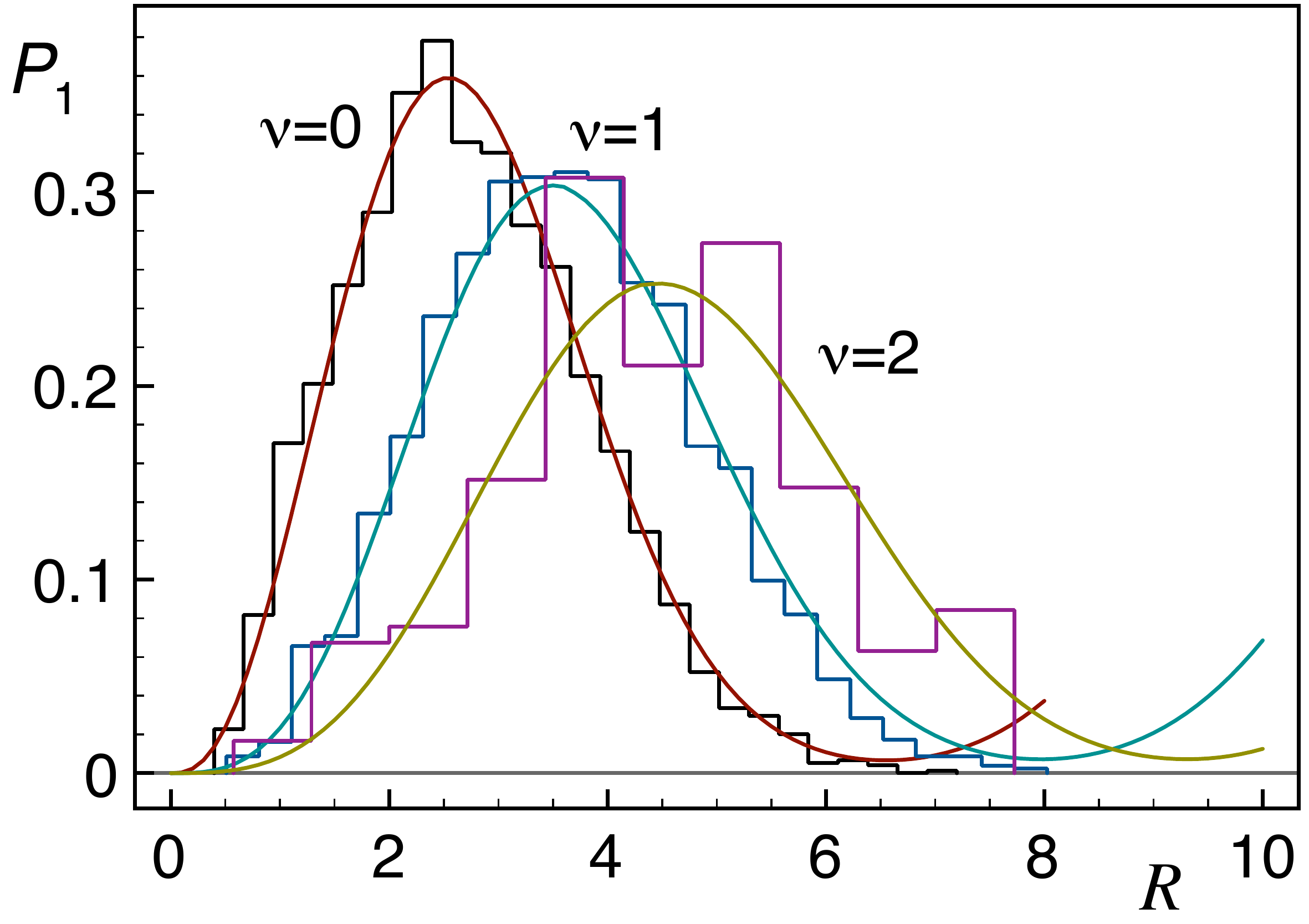}\hspace*{2mm} 
  \includegraphics[width=48mm]{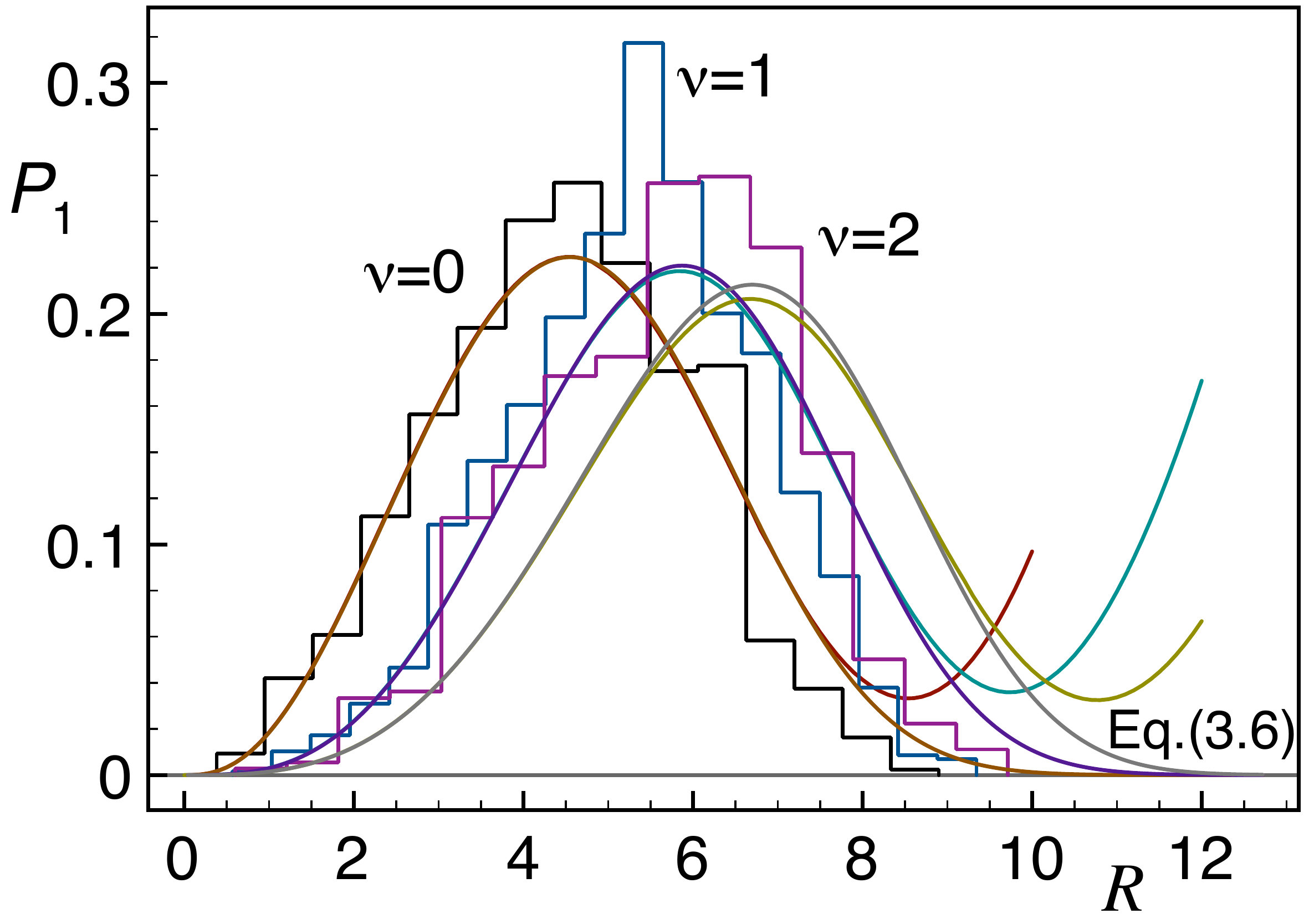} 
  \caption{Integrated distribution $P_1(R)$ of the first eigenvalue
    for $\nu=0$, 1, 2 and $\mu a=0.1$ (left), $\mu a=0.3$ (middle),
    and $\mu a= 1.0$ (right).  The solid lines are the RMT results
    from Eq.~\protect\eqref{p1exp}, the the histograms are the lattice
    data of Ref.~\cite{Bloch:2006cd}. The bending-up of the RMT curves
    for large $R$ is an artifact of truncating the expansion
    \protect\eqref{p1exp}. For $\mu a=1.0$ we also show the exact RMT
    results in the $\alpha\to\infty$ limit from
    Eq.~\protect\eqref{p1}.}
  \label{datamu}
\end{figure}

To compute $p_1$ from the lattice, we choose for our contours
$\partial J[\eta]$ concentric semicircles with radius $R=\eta$ for all
values of $\mu$.  (Other choices are also possible.)  The localized
nature of the IEDs allows us to integrate over the phase (i.e., we
compute $P_1(R)=\int_0^\pi d\theta\, R\,p_1(R,\theta)$) rather than to
consider cuts as in Ref.~\cite{Bloch:2006cd}.  This procedure results
in a much better signal. As a consequence, we are able to obtain a
better comparison in topological sectors $\nu = 0$, 1 and, for the
first time, to successfully test the RMT predictions in the $\nu = 2$
sector. In Fig.~\ref{datamu} we show the comparison of RMT predictions
and lattice results for $\mu a= 0.1$, 0.3, 1.0 and $\nu = 0$, 1, 2
(the $\mu a= 0.2$ case is similar \cite{ABSW}).  We see that for
smaller $\mu$, the agreement is excellent, whereas there are
deviations for $\mu a= 1.0$ and $\nu = 1$, 2.  In these cases we are
outside the $\epsilon$-regime of QCD so that RMT no longer applies.
We emphasize that while the ascent of the distributions from zero was
in principle already tested in Ref.~\cite{Bloch:2006cd} through the
density, their descent represents a new, parameter-free test.

The deviations of the theoretical curves from zero for large $R$ are
an artifact of the truncation of the Fredholm determinant expansion.
As $\mu$ (or $\alpha$) is increased, the convergence of the
approximation becomes slower, i.e., more terms are needed.  

For $\mu a= 1.0$ we are almost in the radially symmetric regime. Thus
we expect $P_1(R)$ computed approximately through Eq.~\eqref{p1exp} to
be close to the exact result Eq.~\eqref{p1} in the $\alpha\to\infty$
limit.  This expectation is confirmed in Fig.~\ref{datamu} (right).

\section{Conclusions}
\label{concl}

In this work we have studied individual Dirac eigenvalue distributions
in the $\epsilon$-regime of QCD at nonzero chemical potential.  We
provided a general framework for ordering complex eigenvalues.  Our
RMT computation for arbitrary $\alpha$ was based on the truncation of
a Fredholm determinant expansion.  In the $\alpha\to\infty$ limit, we
were able to derive all IEDs analytically in closed form.  These
predictions were then tested against lattice data based on the
generalization of the overlap Dirac operator to nonzero chemical
potential, in the topological sectors $\nu = 0$, 1, 2. We found
excellent agreement between RMT and lattice results for several values
of $\mu$ in the domain of the applicability of RMT. The descent of the
IEDs represents a new, parameter-free test of RMT predictions. The
much improved signal (resulting from the integration over the phase)
allowed us, for the first time, to successfully test the RMT
predictions in the topological sector $\nu = 2$.

\acknowledgments

This work was supported by EPSRC grant EP/D031613/1 (GA \& LS), by EU
network ENRAGE MRTN-CT-2004-005616 (GA) and by DFG grant FOR 465 (JB
\& TW).


\begin{thebibliography}{99}

\bibitem{RMT} 

  E.~V.~Shuryak and J.~J.~M.~Verbaarschot,
  Nucl.\ Phys.\  A {\bf 560} (1993) 306
  [hep-th/9212088];\\
  J.~J.~M.~Verbaarschot,
  Phys.\ Rev.\ Lett.\  {\bf 72} (1994) 2531
  [hep-th/9401059].

\bibitem{ADMN}

  G.~Akemann, P.~H.~Damgaard, U.~Magnea and S.~Nishigaki,
  Nucl.\ Phys.\  B {\bf 487} (1997) 721
  [hep-th/9609174];\\
  P.~H.~Damgaard and S.~M.~Nishigaki,
  Nucl.\ Phys.\  B {\bf 518} (1998) 495
  [hep-th/9711023].

\bibitem{pk_real}

  T.~Wilke, T.~Guhr and T.~Wettig,
  Phys.\ Rev.\  D {\bf 57} (1998) 6486
  [hep-th/9711057];\\
  S.~M.~Nishigaki, P.~H.~Damgaard and T.~Wettig,
  Phys.\ Rev.\  D {\bf 58} (1998) 087704
  [hep-th/9803007];\\
  P.~H.~Damgaard and S.~M.~Nishigaki,
  Phys.\ Rev.\  D {\bf 63} (2001) 045012
  [hep-th/0006111].

\bibitem{James} 

  J.~C.~Osborn,
  Phys.\ Rev.\ Lett.\  {\bf 93} (2004) 222001
  [hep-th/0403131];\\
  G.~Akemann, J.~C.~Osborn, K.~Splittorff and J.~J.~M.~Verbaarschot,
  Nucl.\ Phys.\  B {\bf 712} (2005) 287
  [hep-th/0411030].

\bibitem{Misha} 

  M.~A.~Stephanov,
  Phys.\ Rev.\ Lett.\  {\bf 76} (1996) 4472
  [hep-lat/9604003].

\bibitem{SplitV}

  K.~Splittorff and J.~J.~M.~Verbaarschot,
  Nucl.\ Phys.\  B {\bf 683} (2004) 467
  [hep-th/0310271].

\bibitem{BA}

  F.~Basile and G.~Akemann,
  arXiv:0710.0376 [hep-th].

\bibitem{AOW} 

  G.~Akemann and T.~Wettig,
  Phys.\ Rev.\ Lett.\  {\bf 92} (2004) 102002
  [Erratum-ibid.\  {\bf 96} (2006) 029902]
  [hep-lat/0308003];\\
  J.~C.~Osborn and T.~Wettig,
  PoS {\bf LAT2005} (2006) 200
  [hep-lat/0510115].

\bibitem{Bloch:2006cd}

  J.~Bloch and T.~Wettig,
  Phys.\ Rev.\ Lett.\  {\bf 97} (2006) 012003
  [hep-lat/0604020].

\bibitem{Toublan:1999hx}

  D.~Toublan and J.~J.~M.~Verbaarschot,
  Int.\ J.\ Mod.\ Phys.\  B {\bf 15} (2001) 1404
  [hep-th/0001110].

\bibitem{GHS88} 

  R.~Grobe, F.~Haake and H.-J.~Sommers,
  Phys.\ Rev.\ Lett.\ {\bf 61} (1988) 1899.

\bibitem{Markum:1999yr}

  H.~Markum, R.~Pullirsch and T.~Wettig,
  Phys.\ Rev.\ Lett.\  {\bf 83} (1999) 484
  [hep-lat/9906020].

\bibitem{Akemann:2003tv}

  G.~Akemann and P.~H.~Damgaard,
  Phys.\ Lett.\  B {\bf 583} (2004) 199
  [hep-th/0311171].

\bibitem{AS}

  G.~Akemann and L.~Shifrin (2007), to be published.

\bibitem{Bloch:2007xi}

  J.~Bloch and T.~Wettig,
  arXiv:0709.4630 [hep-lat],
  to appear in Phys.\ Rev.\ D.

\bibitem{ABSW}

  G.~Akemann, J.~C.~R.~Bloch, L.~Shifrin and T.~Wettig,
  arXiv:0710.2865 [hep-lat].

\bibitem{overlap} 

  R.~Narayanan and H.~Neuberger,
  Nucl.\ Phys.\  B {\bf 443} (1995) 305
  [hep-th/9411108];\\
  H.~Neuberger,
  Phys.\ Lett.\  B {\bf 417} (1998) 141
  [hep-lat/9707022].

\end{thebibliography}
\end{document}